\documentclass[
    amsmath,
    amssymb,
    aps,
    floatfix,
    twocolumn
]{revtex4-2}
\usepackage[utf8]{inputenc}
\usepackage[colorlinks=true,citecolor=blue,allcolors=blue]{hyperref}
\usepackage{amsmath,amsthm,amsfonts,amssymb,amscd}
\usepackage{graphicx}
\usepackage{lastpage}
\usepackage{enumerate}
\usepackage{mathrsfs}
\usepackage{bm}
\usepackage{braket}
\usepackage[version=4]{mhchem}
\usepackage{blindtext}
\usepackage{natbib}

\newcommand\templangle{\langle}
\newcommand\temprangle{\rangle}
\newcommand\mati{\begin{matrix}}
\newcommand\matf{\end{matrix}}
\newcommand\bmati{\begin{bmatrix}}
\newcommand\bmatf{\end{bmatrix}}
\newcommand\pmati{\begin{pmatrix}}
\newcommand\pmatf{\end{pmatrix}}

\newcommand{\pmat}[1]{\pmati #1 \pmatf}

\newcommand{\abs}[1]{\left\lvert#1\right\rvert}

\newcommand{\pushright}[1]{\ifmeasuring@#1\else\omit\hfill$\displaystyle#1$\fi\ignorespaces}
\newcommand{\pushleft}[1]{\ifmeasuring@#1\else\omit$\displaystyle#1$\hfill\fi\ignorespaces}

\newcommand{\quotes}[1]{``#1''}

\newcommand{\pket}[1]{\renewcommand{\rangle}{)} \ket{#1} \renewcommand{\rangle}{\temprangle}}

\newcommand{\pbraket}[1]{\renewcommand{\langle}{(}\renewcommand{\rangle}{)} \braket{#1} \renewcommand{\rangle}{\temprangle}\renewcommand{\langle}{\templangle}}

\begin{document}

\author{
    Joshua Forer$^{1,2}$,
    D\'avid Hvizdo\v{s}$^3$,
    Xianwu Jiang$^4$,
    Mehdi Ayouz$^5$,
    Chris H. Greene$^3$,
    Viatcheslav Kokoouline$^1$
}

\homepage{E-mail: slavako@ucf.edu}
\affiliation{
    $^1$Department of Physics, University of Central Florida, 32816, Florida, USA \\
    \mbox{$^2$Institut des Sciences Moléculaires, Université de Bordeaux, CNRS UMR 5255, 33405, Talence Cedex, France} \\
    $^3$Department of Physics and Astronomy and
    Purdue Quantum Science and Engineering Institute, Purdue University, West Lafayette, Indiana 47907, USA \\
    $^4$Department of Physics, Wuhan University of Technology, Wuhan 430074, China\\
    $^5$Université Paris-Saclay, CentraleSupélec, Laboratoire de Génie des Procédés et Matériaux, \\
    91190, Gif-sur-Yvette, France. \\
    \looseness=-3 % avoid linesbreaks in affiliations
}

\title{Unified treatment of resonant and non-resonant mechanisms in dissociative recombination: benchmark study of \ce{CH+}
}

\date{\today}% It is always \today, today,

\begin{abstract}
The theoretical approach developed here treats uniformly the direct and indirect mechanisms of dissociative recombination (DR) in a diatomic ion.
The present theory is based on electron scattering calculations performed at several internuclear distances in the molecule.
It is easy to implement because there is no need to separately evaluate couplings and the bound dissociative states of the neutral molecule.
The theory can be applied to molecular ions with or without electronic resonances at low energies.
The approach is applied to compute the DR cross section in electron-CH$^+$ collisions.
The computed cross section agrees generally well with recent state-resolved data from a cryogenic storage-experiment, which validates the approach.
\end{abstract}

\maketitle

%\tableofcontents

\section{Motivation}
Dissociative recombination plays an important role in molecular plasma environments, such as combustion, plasma-based technologies, planetary atmospheres, the interstellar medium, and other fields.
Cross sections for this process are needed for understanding, modelling, predicting, and controlling (in technological applications, for example) the plasma.
Accurate cross sections can be obtained experimentally in storage rings or afterglow plasma experiments \cite{larsson1997dissociative,florescu06,adams2006electron,larsson2008dissociative}.
However, the experimental approach is expensive and not suitable for reactive species and molecular ions in an excited state (rotational, vibrational, or electronic).
On the other hand, a theoretical approach with now-available abundant computational resources can provide the data for the situations where the experiment is difficult or impossible.

Theory based on first principles has only recently reached the point where it can, at least in principle, accurately describe the DR process accounting for all degrees of freedom: electronic, vibrational, and rotational.
A recently demonstrated success was for the benchmark system of HeH$^+$ \cite{vcurik2020dissociative,vcurik2017inelastic}: the theory reproduces all details of the experimental DR cross sections obtained in the cryogenic storage ring (CSR).
But for diatomic ions with more complex electronic structure, the theory is still not as accurate as for HeH$^+$.
However, for the molecular ions having no electronic resonances at low collision energies (i.e., dominated by the indirect DR mechanism), the theoretical approach based on quantum defect theory \cite{hamilton2002competition,kokoouline2003unified,vcurik2008rates,vcurik2020dissociative,vcurik2017inelastic} provides a satisfactory description of the process.
For molecular ions having one or a few electronic resonances at low scattering energies (dominated by the direct DR mechanism), the time-dependent approach by Orel {et al.} \cite{orel93,orel00,larson05b,morisset07} and time-independent approaches developed by Giusti, Takagi, Guberman, and Schneider { et al.} \cite{giusti1977dissociative,giusti1980multichannel,guberman1991generation,takagi1991dissociative,guberman1995dissociative,guberman1997mechanism,guberman2007role,takagi2013cross,carata2000core,chakrabarti2018dissociative} are able to represent the process and provide rate coefficients that generally agree with experimental results, with the exception of certain molecular ions, such as CH$^+$.
The quantum defect theories, the time-dependent, and the time-independent approaches are not well adapted for treating DR of an ion having many electronic resonances at low scattering energies.
This is the case, for example, for the ions having one or several excited states at relatively low energies, 1-3 eV.
In this situation, such a low excited state generates a Rydberg series of electronic resonances, which could be excited by a low-energy electron.
Our recent study \cite{jiang2021theory}, proposes a general approach that is able to describe that situation.
Additionally, the approach can be used for ions in which the direct and indirect mechanisms are competing.
In the previous study, the rotational structure of the target ion was not accounted for, i.e. the approach can be used in applications where rotational structure is not important.
In that study, the fixed-nuclei scattering matrix was evaluated at a relatively high scattering energy, 3-5 eV (depending on the internuclear distance), which means that it may not be accurate to represent the scattering at very low energies, say, below 1~eV.
In the present study, we extend the approach to account for the rotational structure of the molecule and also use the fixed-nuclei scattering matrix obtained for low scattering energies.
The approach is applied to the CH$^+$ molecular ion, a good example of a system having low-energy excited states with available experimental data from two different storage ring experiments.
\begin{figure}
	\includegraphics[scale=0.28]{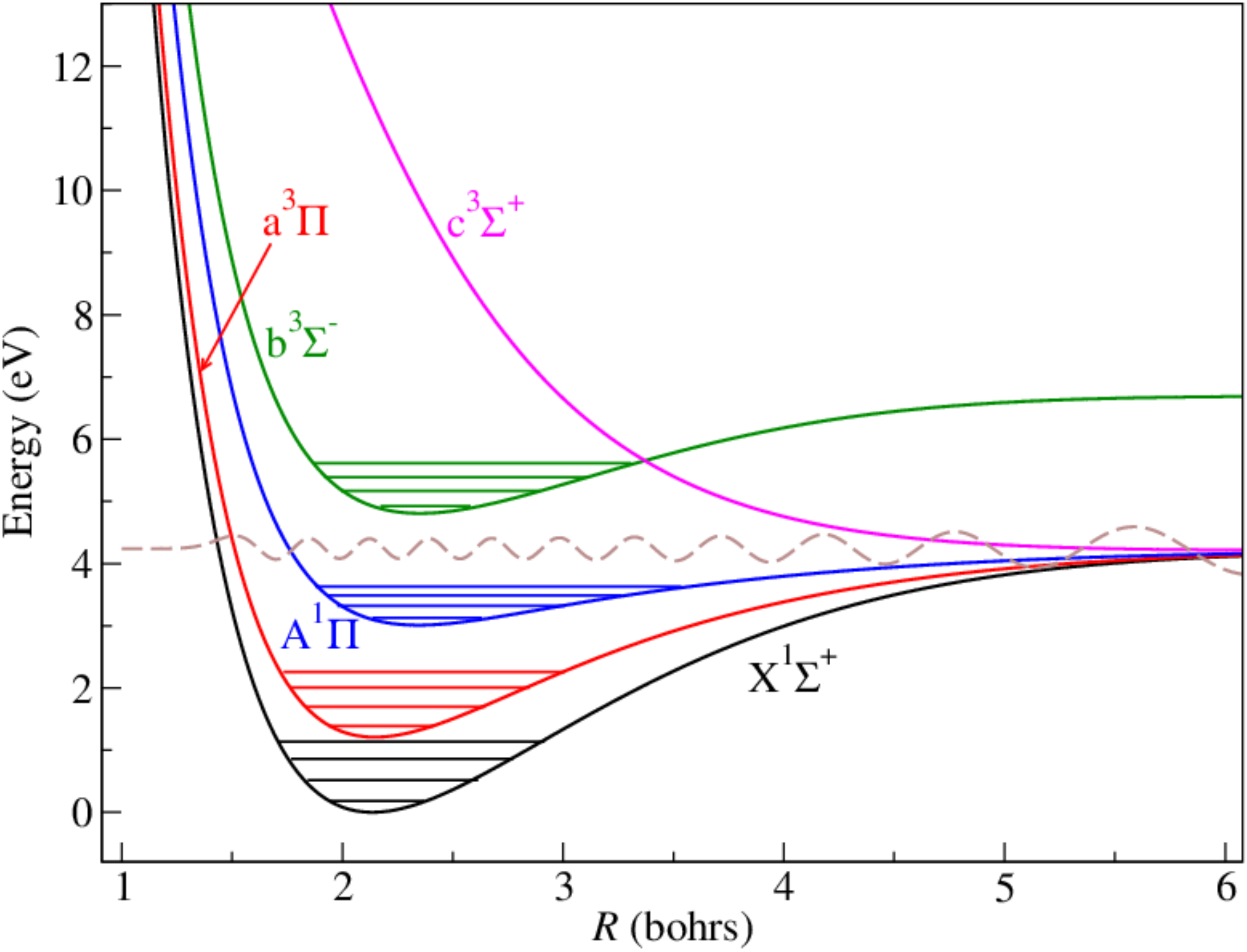}
	\caption{Lowest potential energy curves of the CH$^+$ ion. The dashed line shows the wave function of one of the dissociative vibrational states, obtained with the CAP.
	}
	\label{fig:pes}
\end{figure}

\section{Fixed-nuclei electron scattering matrix}\;
The lowest potential energy curves of the CH$^+$ ion are shown in Fig.~\ref{fig:pes}.
In the storage-ring and low-temperature plasma experiments, the target ion is in the ground electronic state, X$^1\Sigma^+$.
The lowest excited  a$^3\Pi$ and A$^1\Pi$ states of the ion have relatively low energies, 1-3 eV above X$^1\Sigma^+$ and, therefore, generate Rydberg series of electronic resonances, which can be excited in low-energy collisions.
These resonances have to be accounted for in the model describing the electron scattering from the molecular ion.
In particular, the electronic resonances provide pathways for a significant probability of electron capture into dissociating states of CH and, therefore, the process of dissociative recombination.
As in the previous study \cite{jiang2021theory}, the electron scattering calculations at fixed geometries were performed using the UK R-matrix code  \cite{tennyson2010electron} and Quantemol-N suite \cite{tennyson2007quantemol}.

To perform the scattering calculations, the configuration-interaction method (CI) was employed: as a basis to represent the electronic state of the target ion, the correlation consistent polarized valence quadruple-zeta (cc-pVQZ) basis set was used.
In the CI expansion, the lowest orbital $1\sigma^2$ was frozen, while the remaining four electrons were distributed over the  $2-7\sigma, 1-3\pi, 1\delta$ orbitals of the active space.
Three $\sigma$ and two $\pi$ virtual orbitals were used to augment the continuum orbital set in the inner region scattering wave function.
The three lowest target states X$^1\Sigma^+$, a$^3\Pi$, and A$^1\Pi$ were retained in the closed-coupling expansion in the outer region.
The R-matrix radius was 13 bohr radii.
The calculations were performed for irreps of the $C_{2v}$ abelian group, for 300 values of the internuclear distance $R$ in the interval from 1.200 to 2.695 bohr with a constant grid step.
The electronic scattering calculations, as well as the cross section calculations, discussed below, were performed in the coordinate system with the origin at the center of mass CH$^+$.
The present treatment differs significantly from the previous study \cite{jiang2021theory} in the way the reactance matrix $\hat{K}$ is obtained. In the previous study, the K-matrix was obtained by performing the scattering calculations at energies above the second excited state A$^1\Pi$ of the ion, where the three lowest electronic states of CH$^+$ are all open for ionization.
The K-matrix obtained at such relatively high energies (3-5 eV) does not have resonances associated with the closed electronic channels (a$^3\Pi$ or A$^1\Pi$) and is used, after the rovibronic channel elimination procedure, to compute the DR cross sections at much lower energies.
However, such a K-matrix is not expected to accurately represent the scattering at low energies.

Here, this problem is solved as follows: the UK R-matrix code is used to produce the full energy-dependent R-matrix in the Wigner-Eisenbud-type form \cite{wigner1947higher, tennyson2010electron}, at the R-matrix box radius (13 bohr) for some small energy $E_{rmat}$, 0.075~eV above the ground electronic state of CH$^+$ for each value of $R$.
When the reactance matrix $\hat{K}$ is computed from the R-matrix, some channels associated with the first excited electronic state (a$^3\Pi$) of CH$^+$, are treated as being open even if they are closed asymptotically for 0.075~eV.
In the language of multichannel quantum defect theory (MQDT), these are usually referred to as {\it weakly closed channels}.
The K-matrix obtained in this way does not exhibit electronic Rydberg resonances produced by the excited states of the ion, although that physics is implicitly contained in $\hat{K}$, and it describes more accurately the scattering of the electron after the MQDT closed-channel elimination procedure is applied.

\begin{figure}
	\centering
	\includegraphics[page=1,width=0.49\textwidth]{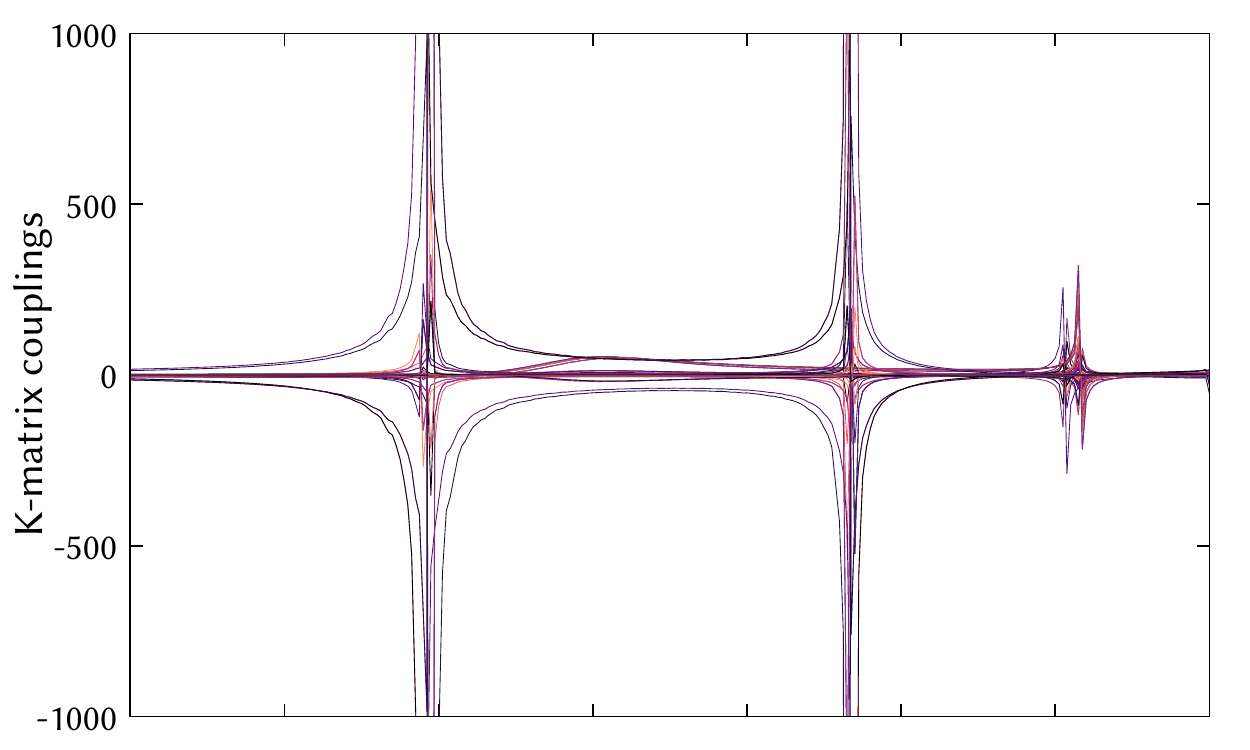}\\
        \vspace{-1em}
        \includegraphics[page=2,width=0.49\textwidth]{fig2.pdf}
  \caption{
    Elements of the K-matrix and scattering-phase matrix as functions of the internuclear distance.
    The notation for each coupling's partial waves is of the form $l'\lambda'\sim l\lambda$.
    The plotted matrix elements labeled in (b) describe couplings between channels in the same electronic target state in the C$_{\infty v}$ point group.
  }
	\label{fig:K_R}
\end{figure}

%With the 3 electronic states  of the ion, two of which are doubly degenerate (a$^3\Pi$ and A$^1\Pi$), and 9 partial waves $l=0-2$ of the incident electron (only channels attached to the a$^3\Pi$ are treated as weakly closed), the calculation yields a {fixed-$R$} K-matrix in a representation of 27 scattering channels.
Scattering channels with partial waves $l = 0 - 2$ for the incoming electron attached to the X$^1\Sigma^+$ and doubly degenerate a$^3\Pi$ state are considered open or weakly closed, resulting in a fixed-$R$ K-matrix in a representation of 27 scattering channels.
Channels attached to the A$^1\Pi$ state are treated as strongly closed.
The K-matrix elements exhibit strong discontinuities as a function of $R$ due to divergence when scattering eigenphase shifts approach $\pi/2$ (modulo $\pi$).
Figure~\ref{fig:K_R} demonstrates this for a few elements of the K-matrix as a function of $R$.
If the vibrational frame transformation would be implemented directly for the K-matrix, which would require an integration over $R$, the strong $R$-dependence of the K-matrix would introduce inaccuracy.
To circumvent this issue, we construct the matrix $\hat{\delta}(R)$ of scattering phase shifts from the K-matrix.
The two matrices are related formally to each other as
\begin{equation}
	\hat{K}(R) = \tan \hat{\delta}(R)\,.
\end{equation}
In practice, the matrix $\hat{\delta}(R)$ in the above equation is obtained in three steps. (1) The matrix $\hat{K}(R)$ is diagonalized for each value of $R$:
\begin{equation}
	\hat{K}(R) = U  K^e(R)U^\dagger\,,
	\label{eq:K_eigen}
\end{equation}
where $\hat{K}^e(R)$ is a diagonal matrix of eigenvalues $K^e_i(R)$ of $\hat{K}(R)$.
(2) The eigenvalues are transformed to eigenphases  $\delta^e_i(R)=\arctan[K^e_i(R)]$ with continuity ensured through appropriate branch choices.
(3) Finally, the matrix  $\hat{\delta}(R)$ is obtained by the same transformation as Eq.~(\ref{eq:K_eigen}),
\begin{equation}
	\hat{\delta}(R) = U \hat\delta^e(R)U^\dagger\,,
\end{equation}
where $\hat\delta^e(R)$ is the diagonal matrix of the eigenphases.

Then the rovibrational frame transformation can be carried out directly on the smoother phase matrix, analogous to the quantum defect matrix frame transformation method that was introduced in Ref.~\cite{DuGreene1986a}.
The transformation for the incident electron channel functions from real-valued  $Y_{l\lambda}$ to ordinary complex-valued ones $Y_l^\lambda$ is
\begin{equation}
	\pmat{ Y_l^{\abs{\lambda}}\hfill \\ Y_l^{-\abs{\lambda}} }
	=
	\frac{1}{\sqrt{2}} \pmat{
		(-1)^\lambda & i(-1)^\lambda \\
		1            & -i
	}
	\pmat{ Y_{l\abs{\lambda}} \hfill \\ Y_{l-\abs{\lambda}} }\,,
\end{equation}
where $\lambda$ is the projection of angular momentum of the electron on the molecular axis.
The transformation for the electronic states of the target ion is
\begin{equation}
	\pmat{\Pi^+ \\ \Pi^-}
	=
	\frac{1}{\sqrt{2}}\pmat{ -1 & -i \\ 1 & -i }
	\pmat{ \Pi_{B_1} \\ \Pi_{B_2} }\,,
\end{equation}
where indices $B_1$ and $B_2$ refer to the two degenerate components of the $^1\Pi$ or $^3\Pi$ states, transforming in the $C_{2v}$ group according to the corresponding irreducible representations.

The fixed-$R$ S-matrix is obtained from the phase-shift matrix using the eigenvalue basis via the following relation
\begin{equation}
\label{eq:S_R}
	\hat{S}^\Lambda(R) = e^{2i\hat{\delta}^\Lambda(R)}\,,
\end{equation}
which is unitary and symmetric by construction and also block-diagonal with respect to $\Lambda$.

%%%%%%%%%%%%%%%%%%%%%%%%%%%%%%%%%%%%%%%%%%%%%%%%%%%%%%%%%%%%%%%%%%%%%%%%%%%%%%%%%%%%%%%%%%%%%%%%%%%%%%%%%%%%%%%%%%%

\section{Rovibronic frame transformations}
The $R$-dependent scattering matrix $\hat{S}^\Lambda(R)$ is transformed into the MQDT scattering matrix with the rovibronic channel indices of CH$^+$ using the rovibronic frame transformation.
The rovibronic channels for any chosen angular momentum and parity are enumerated with the quantum number $n=1-5$ labeling the electronic state of the ion, i.e. one of the five states (X$^1\Sigma^+$, a$^3\Pi$, and A$^1\Pi$) including the (non-spin) degeneracies, with the vibrational quantum number $v$, the angular momentum of the ion $j$, and its projection $\mu$ of the molecular axis.
The transformation is made in two stages, vibrational and rotational.
As in previous studies \cite{hamilton2002competition,kokoouline05}, for the vibrational frame transformation, bound and continuous-spectrum vibrational states of the ion are needed.
The continous-spectrum states are chosen with outgoing-wave boundary conditions to represent outgoing dissociative flux.
Such states are obtained by solving the vibrational Schr\"odinger equation with the Hamiltonian including a complex absorbing potential (CAP) \cite{vibok1992} for each electronic state ($n$) of \ce{CH+}.
The CAP is nonzero only outside of the interaction region, for $R>R_0$, where it has the form
\begin{equation}
V_\textrm{CAP}=	- i\eta N e^{-\frac{2L}{R-R_0}}\,.
	\label{hamiltonian}
\end{equation}
Here $N$ is a normalization constant equal to 13.22 \cite{vibok1992}, $\eta$ is the CAP strength (0.06 atomic units), $L$ is the CAP length (1.2 bohr), and $R_0$ is where the CAP begins (4.8 bohr).
\begin{equation}
	\pbraket{n'v'|nv}
	=
	\int \text{d}R \; \phi_{n'v'}(R) \; \phi_{nv}(R)
	=
	\delta_{n',n} \; \delta_{v',v}\,.
\end{equation}
Due to the unusual properties of the vibrational scalar product for the states $ \phi_{nv}(R)$ obtained with the CAP, without complex conjugating the vibrational part of the complete channel wave functions, in the space of eigenstates $ \phi_{nv}(R)$, the formulas of unitarity of the scattering matrix are slightly different from those for the states with zero-boundary conditions.
The fixed-$R$ S-matrix of Eq.~(\ref{eq:S_R}) is unitary, i.e. matrix product of its Hermitian-conjugate with the matrix itself gives an identity matrix.
Physically, it means that the summed probabilities of scattering (at a fixed $R$) from a given channel to all other possible channels is unity.
The scattering matrix in the basis of vibrational states, called hereafter $\hat{S}^{\Lambda}$, i.e. after the vibrational frame transformation, should also be unitary, if we consider the frame transformation as a unitary operation.  %{\bf Slava: Could we try to talk about this some day, perhaps?  This does not seem to fit the math definition of a unitary matrix.  Do you regard this as a generalized type of unitary matrix?  I think you are citing our 2005 PRA article for this development, but I'm not sure whether you are handling the frame transformation exactly the same as in that paper, are you? We don't have to discuss this before resubmitting the article, of course.}
One can show that the matrix product $\hat{S}^{\Lambda\ddagger}\hat{S}^{\Lambda} $ gives an identity matrix, where elements of matrices in the product are computed as
\begin{align}
	S^{\Lambda}_{n'v'l'\lambda',nvl\lambda}
	&=
	\int \text{d}R \; \phi_{n'v'}(R) \; S^\Lambda_{n'l'\lambda',nl\lambda}(R) \; \phi_{nv}(R)\,\notag
\end{align}
and
\begin{align}
	S^{\Lambda\ddagger}_{nvl\lambda,n'v'l'\lambda'}
	&=
	\int \text{d}R \; \phi_{nv}(R) \; S^{\Lambda *}_{n'l'\lambda',nl\lambda}(R) \; \phi_{n'v'}(R)\,.\notag
\end{align}
Because the S-matrix elements before the frame transformation are complex, but the vibrational states on the left in the integrals are not complex conjugated, the two matrices $S^{\Lambda\ddagger}$ and $S^{\Lambda} $  are not Hermitian conjugate to each other.
On the other hand, submatrices of $S^{\Lambda\ddagger}$ and $S^{\Lambda} $, corresponding to the set of all bound vibrational states (with zero boundary conditions), which are real-valued, are Hermitian conjugate to each other.
Therefore, the generalized unitarity of the scattering matrix with the states with outgoing-wave boundary conditions is consistent with previous theories of frame transformation.
After the vibrational frame transformation, the rotational frame transformation is applied as follows:
\begin{widetext}
	\begin{align}
    S^{J}_{n'v'j'\mu' l',nvj\mu l}
		&=
		\sum_\Lambda \sum_{\lambda'\lambda} (-1)^{l'+\lambda'+l+\lambda} \;
		C^{j'\mu'}_{l'-\lambda',J\Lambda} \; S^{\Lambda}_{n'v'l'\lambda',nvl\lambda} \; C^{j\mu}_{l-\lambda,J\Lambda}
		\\
    S^{J\ddagger}_{nvj\mu l,n'v'j'\mu' l'}
		&=
		\sum_\Lambda \sum_{\lambda'\lambda} (-1)^{l'+\lambda'+l+\lambda} \;
		C^{j\mu}_{l-\lambda,J\Lambda} \; S^{\Lambda\ddagger}_{nvl\lambda,n'v'l'\lambda'} \; C^{j'\mu'}_{l'-\lambda',J\Lambda}\,,
	\end{align}
\end{widetext}
where $\vec{J}=\vec{j}+\vec{l}$ is the total angular momentum of the $\ce{CH+} + \ce{e-}$ system.
Here, it should be stressed that the rotational frame transformation is performed using the standard scalar product rules, with the complex conjugation because the rotational functions are normalized with the standard scalar product.
This transformation is also unitary, i.e. matrix product of the two S-matrices is an identity matrix.
The rovibronically resolved S-matrices, $\hat{S}^{J}$ and $\hat{S}^{J\ddagger}$, are block-diagonal with respect to $J$ and total parity, of course.

%------------------------------------------------------------------------------------------------------------------
\section{Cross section of dissociative recombination}
The total energy $E_\textrm{tot}$ of the $\ce{CH+} + \ce{e-}$ system is the sum of a given collision energy $E_{\textrm{el}}$ and the energy of the initial state of the ion $E_{nvj\mu}$ .
As in previous studies, when some of the channels are closed asymptotically for ionization, the corresponding scattering matrix is obtained from the one $\hat{S}^{J}$ discussed above using  Seaton's channel-elimination procedure.
Matrices $\hat{S}^{J}$ and  $\hat{S}^{J\ddagger}$ are partitioned into blocks defined by energetically open and closed channels:
\begin{equation}
  \hat{S}^{J}%(E_\text{tot})
	=
	\pmat{ \hat{S}_{oo} & \hat{S}_{oc} \\ \hat{S}_{co} & \hat{S}_{cc} },
	\qquad
  \hat{S}^{J\ddagger}%(E_\text{tot})
	=
	\pmat{ \hat{S}^\ddagger_{oo} & \hat{S}^\ddagger_{oc} \\ \hat{S}^\ddagger_{co} & \hat{S}^\ddagger_{cc} }\,.
\end{equation}
Channels are sorted with ascending real part of their energies, with $n_o$ open channels and $n_c$ closed channels.
The diagonal $n_c\times n_c$ matrix  $\hat{\beta}$ is defined
\begin{equation}
	\beta_{i_c'i_c}(E_\text{tot}) = \frac{\pi}{\sqrt{2(E_{i_c} - E_\text{tot})}} \hat{\delta}_{i_c'i_c} \label{beta}
\end{equation}
where $E_{i_c}$ is the energy (complex for vibrational continuum states) of a given closed rovibronic channel $\pket{nvj\mu l}$ indexed by $i_c$.
The closed-channel elimination procedure should be applied separately to both matrices $\hat{S}^J$ and $\hat{S}^{J\ddagger}$   as \cite{kokoouline2003unified}
\begin{gather}
  \hat{S}^{J,\textit{phys}}(E_\text{tot})
	=
	\hat{S}_{oo} - \hat{S}_{oc}\left[\hat{S}_{cc} - e^{-2i\beta}\right]^{-1} \hat{S}_{co},
	\\
  \hat{S}^{J,\textit{phys}\ddagger}(E_\text{tot})
	=
	\hat{S}^\ddagger_{oo} - \hat{S}^\ddagger_{oc}\left[\hat{S}^\ddagger_{cc} - e^{2i\beta^*}\right]^{-1} \hat{S}^\ddagger_{co}
	\label{Sphys}
\end{gather}
In equation (\ref{Sphys}), the sign of the real part of the exponential matrix changes between the $\hat{S}^{J,\textit{phys}\ddagger}_v$ and $\hat{S}^{J,\textit{phys}}_v$ matrices.
The probability for DR to take place, if the initial state of the ion is $\pket{nvj\mu}$ and the total angular momentum is $J$:
%\begin{widetext}
	\begin{align}
		&P^J_{nvj\mu}(E_\text{tot})
    =
		\sum_l \big[ 1 -  \nonumber\\
    &\quad \sum_{l'}\sum_{n'v'j'\mu'}  S^{J,\textit{phys}}_{n'v'j'\mu'l',nvj\mu l}(E_\text{tot})\;S^{J,\textit{phys}\ddagger}_{nvj\mu l,n'v'j'\mu'l'}(E_\text{tot}) \big]\,,
	\end{align}
%\end{widetext}
where the sum over $l$ runs through the possible values of $l$ for each value of $J$ and $j$.
The DR cross section for a given $J$ is
\begin{equation}
	\sigma^J_{nvj\mu}(E_\text{el}) = \frac{\pi}{k^2} \frac{2J+1}{2j+1} P^J_{nvj\mu}(E_\text{tot})\,,
\end{equation}
where $k=\sqrt{2m_\text{e}E_\text{el}}$ is the magnitude of the incident electron wave vector.
The total DR cross section starting from the initial state $\pket{nvj\mu}$ is given by the sum over $J$
\begin{equation}
	\sigma_{nvj\mu}(E_\text{el}) = \sum_J \sigma^J_{nvj\mu}(E_\text{el})\,.
\end{equation}

\section{Comparison with experiment}
\begin{figure}
	\includegraphics[width=0.47\textwidth]{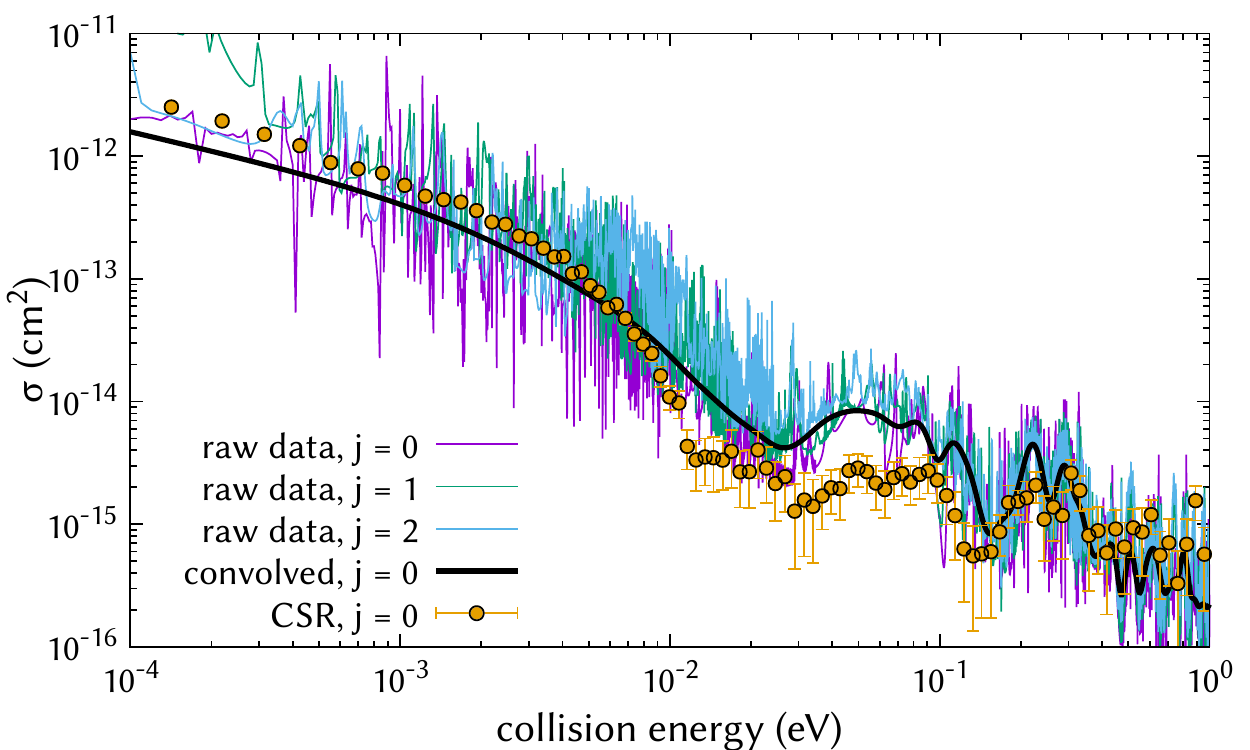}
	\caption{Comparison of the DR cross section obtained in the present study with the recent CSR experiment data \cite{paul_CHp}.
    Three thin solid lines of different color represent cross sections for the CH$^+$ being in the ground vibronic state X$^1\Sigma^+$($v=0$) in three different rotational states $j=0,1,2$.
    The thick solid line represents the $j=0$ theoretical cross section, accounting for uncertainties in collision energy distribution in the experiment.
    The experimental cross section, obtained for $v=0,j=0$, is shown with circles.
	}
	\label{fig:xs_raw}
\end{figure}
Figure \ref{fig:xs_raw} shows the comparison of the DR cross section obtained in the present study with the recent CSR experiment data \cite{paul_CHp}.
The thin solid lines are the theoretical cross sections for the CH$^+$ being in the ground electronic state X$^1\Sigma^+$ and ground vibrational $v=0$ states but in different rotational $j=0,1,2$.
For the $v=0,\ j=0$ experimental data shown in the figure, the abscissa represents the averaged value $E_\parallel$ of the longitudinal collision energy in the experiment, while the cross section values (ordinate) are obtained from the experimental rate coefficients dividing them with velocity corresponding to $E_\parallel$.
The theoretical cross section exhibits many resonances associated with energetically closed rovibronic channels of the ion.
To compare with the experimental data, the theoretical results should account for uncertainties in the collision energies in the experiment.
The electron energy distribution in the experiment is quite complicated and depends on the detuning energy between the electron and ionic beams.
However, constants for longitudinal (0.27~meV) and transverse (2~meV) uncertainties \cite{paul_CHp} and the convolution procedure, employed in previous studies \cite{kokoouline2003unified,santos07}, should represent the major effect of the distribution.
The same convolution procedure as in the previous studies \cite{kokoouline2003unified,santos07} is employed here.
The thick solid line represents the theoretical data convolved with the longitudinal and transverse electron-energy uncertainties for the $j=0$ cross section.
The abscissa and ordinate of the convoluted theoretical cross section have the same meaning as in the shown experimental data and, therefore, can be directly compared to the data.

The convolved theoretical cross section agrees with the experimental cross section generally well, especially when compared with the previous theoretical results.
The next figure, Fig.~\ref{fig:xs_reduced}, compares the present convolved cross section with all available previous experimental and theoretical data.
For a more detailed comparison, the ordinate axis shows the cross section multiplied with energy so that the data does not span several orders of magnitude as in Fig.~\ref{fig:xs_raw}.
In addition to the CSR experimental data \cite{paul_CHp}, there is a previous room-temperature storage ring experiment \cite{amitay1996dissociative} and a single-pass merged-beam experiment \cite{mul1981}.
In these experiments, unlike the CSR, the ions were also hot, at least, rotationally and, most likely, vibrationally in  \cite{mul1981}.
At low energies, the two storage-ring experiments agree with each other and the present theory.
The merged-beam data are somewhat larger, possibly, due to excited vibrational levels that could be present in the ion beam.
Among the previous theoretical results, the best agreement with the experiment at low energies was demonstrated by the study by Takagi \cite{takagi1991dissociative}, where the rotational structure was accounted for, and the study by  \citet{mezei2019dissociative}, where the rotation was not included.

\begin{figure}
	\includegraphics[width=0.45\textwidth]{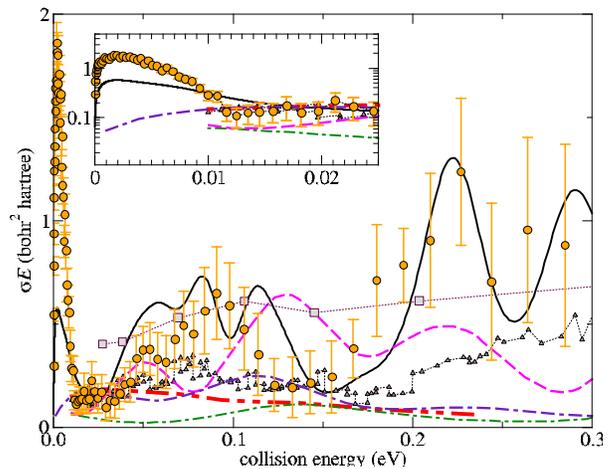}
  \caption{\quotes{Reduced} cross section,  $E_\text{el}\sigma(E_\text{el})$.
    The present theoretical data are shown with the solid (black) line.
    The experimental data are shown with circles \cite{paul_CHp}, triangles \cite{amitay1996dissociative}, squares \cite{mul1981}.
    The previous theoretical data are shown by dashed (magenta) \cite{chakrabarti2018dissociative}, dot-dashed (green) \cite{carata2000core}, double-dot-dashed (red) \cite{takagi1991dissociative}, and dot-double-dashed (indigo)  \cite{mezei2019dissociative} lines.
	}
	\label{fig:xs_reduced}
\end{figure}

\section{Conclusion}
We stress the following major findings and developments reported here in the theory of dissociative recombination.
\begin{itemize}
 \item The theoretical approach, presented in the study, is able to treat the direct and indirect DR mechanisms using the same approach.
   There have been previous studies \cite{takagi1991dissociative,chakrabarti2018dissociative} where the two mechanisms were accounted for at the same time; the present theory appears to be easier to implement because it does not require computing bound dissociative states of the neutral molecule and couplings associated with them.
   In an earlier study of DR in HeH$^+$ \cite{haxton2009ab}, a similar approach was employed.
 \item The theory can be applied to molecular ions with or without electronic resonances at low energies, in particular, to open-shell ions, such as N$_2^+$ or O$_2^+$. The DR process in electron collisions with open-shell ions is particularly difficult to treat using previous theoretical approaches.
 \item The approach can be immediately applied to study rotational, vibrational, and electronic excitations of diatomic ions, including open-shell ones.
   Cross sections for these processes are important for many plasma applications, but only rarely can they be measured in an experiment.
 \item The present cross section of DR in CH$^+$ agrees generally with the recent, most accurate (state-resolved) data from the CSR experiment \cite{paul_CHp}, which validates the approach.
   The theory reproduces not only the overall magnitude of the measured cross sections, but also positions of major resonances resolved in the experiment.
 \item One of the limitations of the present study, which could be improved in the future, is the use of energy-independent body-frame scattering matrices.
   It was recently demonstrated \cite{vcurik2020dissociative} by two of us for the example of  HeH$^+$ DR that accounting for the energy dependence of the scattering matrix improves the agreement with the experiment.
 \item Also, in the present theory, the residual long-range coupling between channels caused by the charge-dipole interaction in the $e^-$-CH$^+$ potential is neglected.
   A future theory should account for the coupling.
\end{itemize}

\section*{Acknowledgements}We are grateful to Oldrich Novotny, Holger Kreckel, and Andreas Wolf from the Max-Planck-Institut f\"ur Kernphysik (MPIK) in Heidelberg, to Abel Kalosi and Daniel Paul, from the MPIK and the Columbia University for providing us with the experimental data as well as for helpful discussions with them and with Ioan F. Schneider (Le Havre) and Zsolt J. Mezei (Debrecen).
This work acknowledges support from the National Science Foundation, Grant Nos.2110279 (UCF) and 2102187 (Purdue), the Fulbright-University of Bordeaux Doctoral Research Award, the Thomas Jefferson Fund of the Office for Science and Technology of the Embassy of France in the United States, and the program \quotes{Accueil des chercheurs étrangers} of CentraleSupélec.

% \bibliography{refs.bib}

\providecommand{\noopsort}[1]{}\providecommand{\singleletter}[1]{#1}%

\end{document}